\documentclass[twocolumn,prl]{revtex4}
\usepackage{graphicx}
\newcommand{\be}{\begin{equation}}
\newcommand{\ee}{\end{equation}}
\newcommand{\bea}{\begin{eqnarray}}
\newcommand{\eea}{\end{eqnarray}}

\newcommand{\la}{\langle}
\newcommand{\ra}{\rangle}

\renewcommand{\phi}{\varphi}
\renewcommand{\epsilon}{\varepsilon}
\renewcommand{\vec}[1]{{\bf #1}}

\begin{document}

\title{Order from Disorder in Graphene Quantum Hall Ferromagnet}
\author{Dmitry A. Abanin, Patrick A. Lee, Leonid S. Levitov}
\affiliation{
 Department of Physics,
 Massachusetts Institute of Technology, 77 Massachusetts Ave,
 Cambridge, MA 02139}

\begin{abstract}
Valley-polarized quantum Hall states in graphene are described
by a Heisenberg O(3) ferromagnet model,
with the ordering type controlled 
by the strength and sign of valley anisotropy. 
A mechanism resulting from electron coupling to strain-induced gauge field,
giving leading contribution to the anisotropy, 
is described in terms of
an effective random magnetic field
aligned with the ferromagnet $z$ axis. We argue that such random field
stabilizes the XY ferromagnet state, which is 
a coherent equal-weight mixture of the $K$ and $K'$ valley states.
The implications such as the Berezinskii-Kosterlitz-Thouless ordering transition 
and topological defects
with half-integer charge are discussed.
\end{abstract}

\maketitle

\begin{figure}
\includegraphics[width=1in]{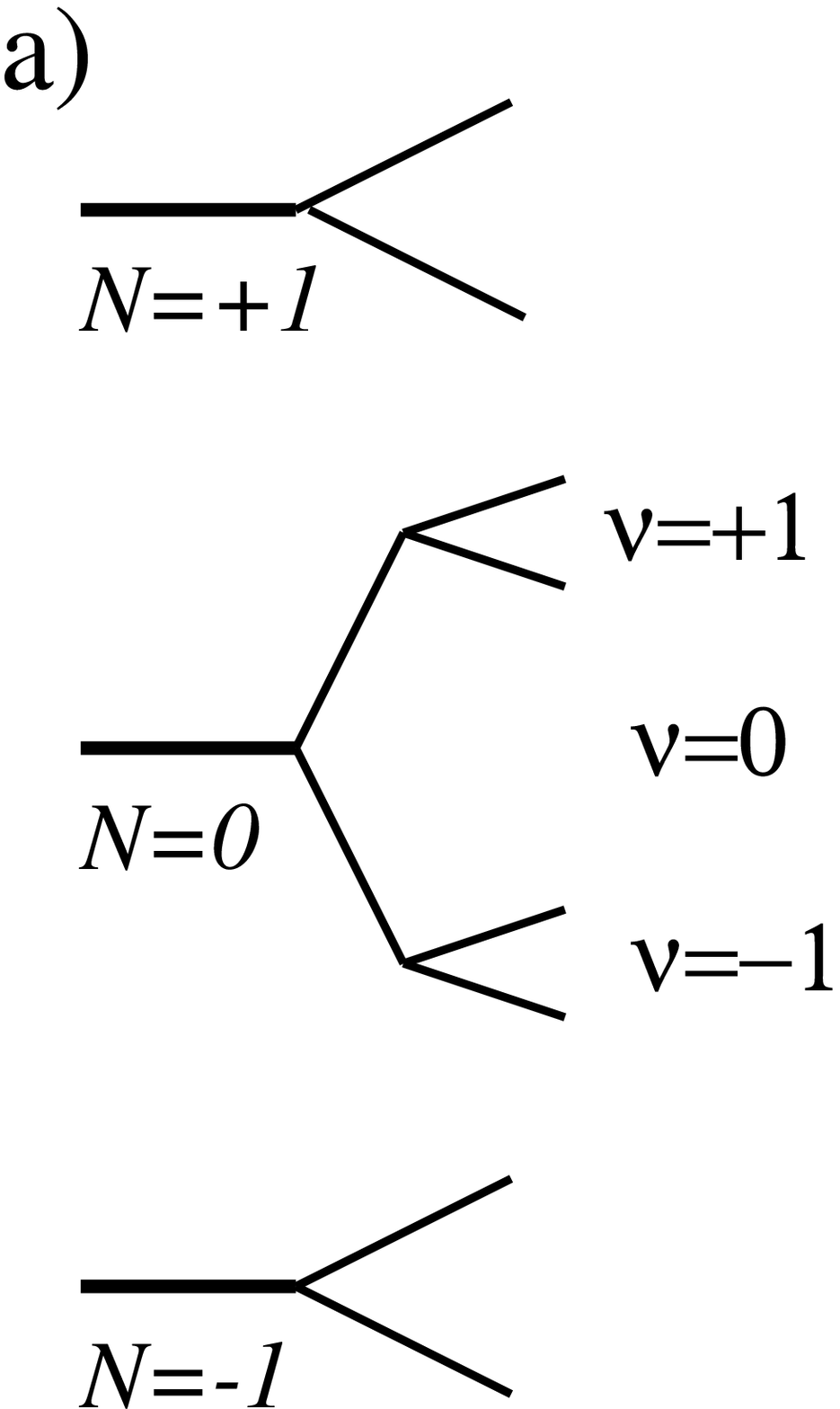}
\includegraphics[width=2.3in]{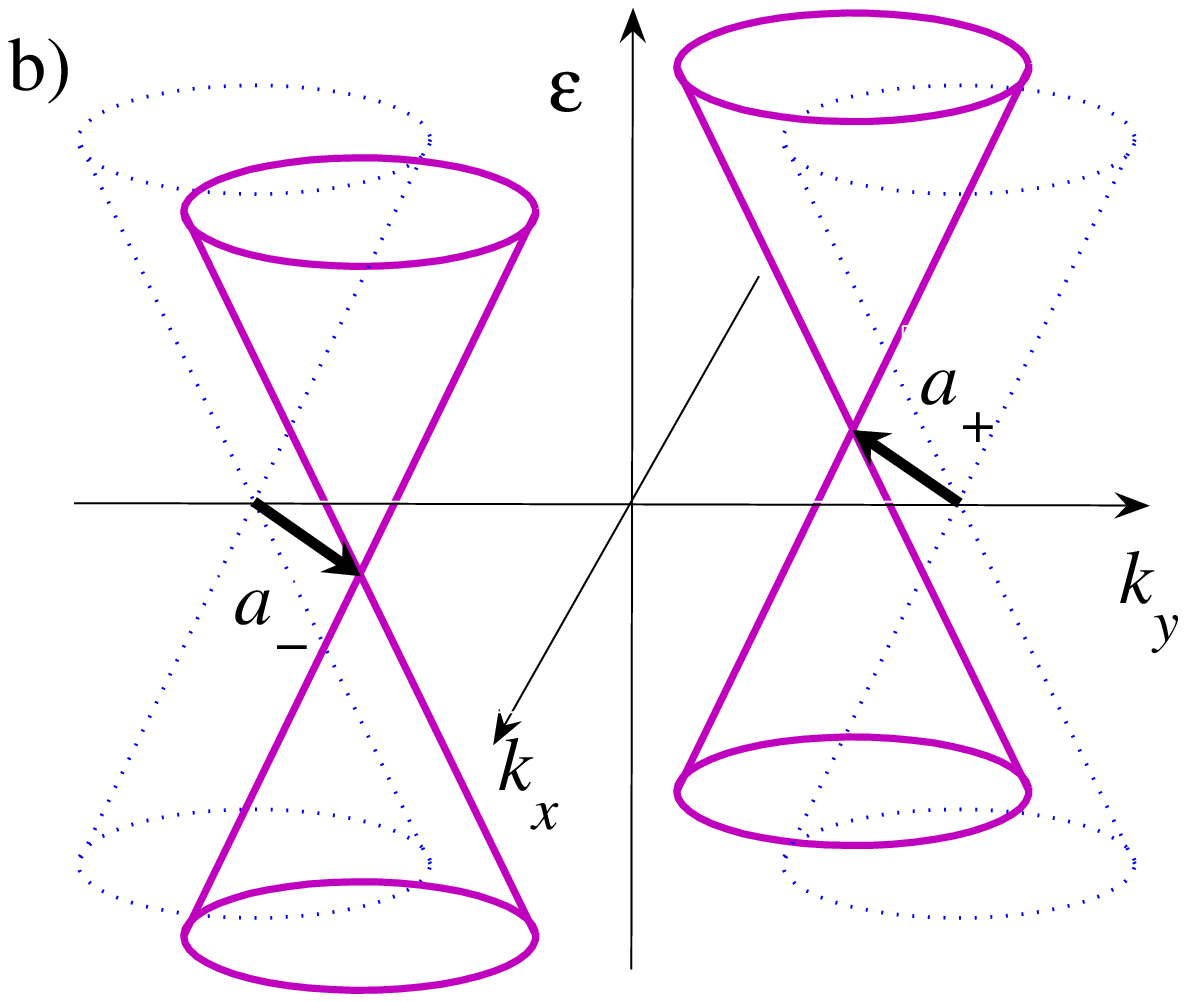}
\vspace{-0.1cm}
\caption[]{ 
a) Graphene Landau level
splitting, Ref.\cite{Zhang06}, attributed to spin and valley polarization.
When the Zeeman energy
exceeds valley anisotropy, 
all $N=0$ states
are spin-polarized, with the $\nu=\pm 1$ states
valley-polarized and the $\nu=0$ state
valley-unpolarized.
b) 
The effect of uniform strain on electron spectrum, 
Ref.\cite{Iordanskii85}, described by 
Dirac cones shift
in opposite directions from the points $K$ and $K'$.
Position-dependent strain is described as 
a random gauge field, Eq.(\ref{eq:dirac_random}).}
\label{fig1}
\end{figure}

Gate-controlled graphene
monolayer sheets\,\cite{Novoselov04} host an interesting 
two-dimensional electron system. 
Recent studies of transport have uncovered,
in particular, anomalous Quantum Hall 
effect\,\cite{Novoselov05,Zhang05}, resulting from Dirac fermion-like behavior
of quasiparticles.
Most recently, when magnetic field
was increased above about 20\,T,
the Landau levels (LL) were found to split\,\cite{Zhang06}, 
with the $n=\pm1$ and $n=0$ levels
forming two and four sub-levels, respectively, as illustrated in Fig.\ref
{fig1}a. 
The observed splittings were attributed to spin and valley degeneracy
lifted by the Zeeman and exchange interactions. 

The physics of the interaction-induced gapped quantum Hall state
is best understood by analogy with the well-studied 
quantum Hall bilayers realized in double quantum well
systems \cite{QHFM}.
In the latter, the interaction is nearly degenerate 
with respect to rotations of pseudospin describing 
the two wells. As a result, the states
with odd filling factors are characterized by pseudospin O(3)
ordering, the so-called quantum Hall ferromagnet 
(QHFM)\,\cite{QHFM}. 
The pseudospin $z$ component describes density imbalance between the wells,
while the $x$ and $y$ components describe the inter-well coherence
of electron states.
Several different phases \cite{MacDonald94,MacDonald95} are possible in QHFM 
depending on the strength of the anisotropic part of Coulomb interaction, 
controlled by well separation.

In the case of graphene, with all electrons moving in a single plane,
the valleys $K$ and $K'$ play the role of the two wells in the pseudospin
representation
with the lattice constant replacing the inter-well separation.
To assess the possibility of QHFM ordering, we note that
the magnetic length at the $10-30\,{\rm T}$ field is 
much greater 
than the lattice constant. 
Thus graphene QHFM can be associated
with the double-well systems with 
nearly perfect pseudospin symmetry of Coulomb interaction \cite{Nomura06,Moessner06,Alicea06}.
Our estimate, presented below, yields anisotropy magnitude
of about $10\,\mu {\rm K}$ at $B\sim
30 \,{\rm T}$, which is very small compared to other energy scales in the 
system. 

Can some other mechanism break pseudospin symmetry more efficiently? 
Coupling to disorder seems an unlikely candidate at first glance.
However, there is an interesting effect that received relatively little 
attention, which is strain-induced random gauge field introduced by 
Iordanskii and Koshelev~\cite{Iordanskii85}. 
To clarify its origin, let us consider the tight-binding model
with spatially varying hopping amplitudes.
Physically, such variation can be due to local strain, 
curvature\,\cite{KaneMele97,Morpurgo06} or chemical
disorder. With hopping amplitudes $t_i$ for three bond orientations 
varying independently, we write
\be\label{eq:dirac}
\left[\begin{array}{cc}
         0 &  \tau ({\bf q})\\
         \tau^*({\bf q}) &  0
      \end{array}
\right] 
\left( 
\begin{array}{c}
        u \\
        v
\end{array}
\right)
 =\epsilon
 \left( 
\begin{array}{c}
        u \\
        v
\end{array}
\right), \,\,\, \tau({\bf q})=\sum_{i=1,2,3} t_i e^{i{\bf q}.{\bf e}_i}, 
\ee
where ${\bf e}_i$ are vectors connecting a lattice site to its 
nearest 
neighbors, and $u$ and $v$ are wavefunction amplitudes 
on the two 
non-equivalent
sublattices, $A$ and $B$. 
The low-energy Hamiltonian
for the valleys $K$ and $K'$ is obtained at
${\bf q}\approx\pm {\bf q}_0$, where $\pm{\bf q}_0$ are two 
non-equivalent Brillouin zone corners:
\be\label{eq:dirac_random}
H_{\pm}=v \left[\begin{array}{cc}
         0 & ip_x\mp p_y+\frac{e}{c}a_{\pm}  \\
         -ip_x \mp p_y+\frac{e}{c}a^*_{\pm} &  0
      \end{array}
\right]
\ee
with $ a_{\pm}=
\frac{c}{e}\sum_{i=1,2,3}\delta t_i e^{\pm i{\bf q}_0. {\bf e}_i}$,
where the subscript $+(-)$ corresponds to $K(K')$ valley. 
Decomposing $a_\pm=a_y\mp ia_x$, we see that the effective vector potential 
in the two valleys is given by $\pm (a_x,a_y)$.  
Notably, the field $a$ is of opposite sign for the two valleys, 
thus preserving 
time-reversal symmetry (see Fig.\ref{fig1}b). 
 
Here we assume that the gauge field 
has white noise correlations with a correlation length $\xi$,
\be\label{Acorr}
\langle a_{i}(k) a_{j}(k) \rangle_{k\xi\ll 1} = \alpha^2, 
\quad a_{i}(k)=\int e^{-ikr}a_{i}(r)d^2r,
\ee
as appropriate for white noise fluctuations of $\delta t_i$.
The fluctuating effective magnetic field can be estimated as
\be\label{eq:delta_h}
\delta h(x)=\partial_x a_y-\partial_y a_x\sim \alpha/\xi^2,
\ee
whereby the correlator of Fourier harmonics $\la \delta h_k\delta h_{-k}\ra$ 
behaves 
as $k^2$ at $k\xi\ll 1$.

Recently, strain-induced effective magnetic field 
was employed to explain anomalously small 
weak localization in graphene\,\cite{Morozov06}.
A direct observation of graphene ripples\,\cite{Morozov06} 
yields typical corrugation length scale $\xi$ of a 
few tens of nanometers. 
Estimates from the first 
principles\,\cite{Morozov06} gave $\delta h\sim 0.1-1\,{\rm T}$, consistent
with the observed degree of weak localization suppression.

\begin{figure}
\includegraphics[width=3.2in]{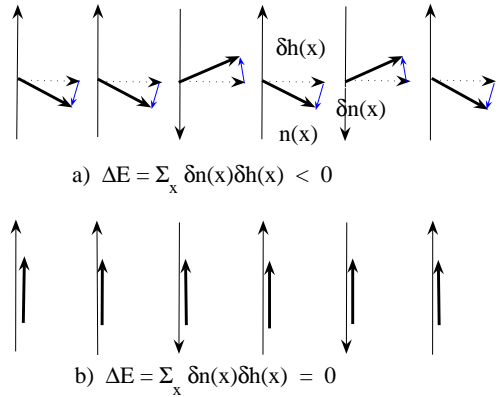}
\vspace{-0.1cm}
\caption[]{
Random field-induced order in a ferromagnet. The energy gained from the
order parameter tilting opposite to the field is maximal when the spins
and the field are perpendicular (a), 
and minimal for the spins parallel to the field (b). 
Uniaxial random field induces
ordering in the transverse plane. 
}
\label{fig2}
\end{figure}

Valley $K$-$K'$ asymmetry of QHFM in the presence 
of the gauge field (\ref{eq:dirac_random})
translates into {\it a uniaxial} random magnetic field, 
proportional to (\ref{eq:delta_h}) and aligned
with the pseudospin $z$ axis. 
We shall see that the effect of random gauge field is subtle: 
somewhat counterintuitively,
weak $\delta h$ induces ordering in the system, acting as an easy plane 
anisotropy which favors the $XY$ state. 
This behavior can be understood by noting that 
the transverse fluctuations in a ferromagnet
are softer than the longitudinal fluctuations, 
making it beneficial for the spins to be 
polarized, on average, transversely
to the field, 
as illustrated 
in Fig.\ref{fig2}.
This {\it random field-induced ordering} maximizes the energy gain
of the spin system coupled to $\delta h$.

For magnets with uniaxial random field this behavior 
has been established\,\cite{Aharony78,Feldman98} in high space dimension. 
The situation in dimension
two is considerably more delicate\,\cite{Pelcovits85,Wehr06} due to 
competition with
the Larkin-Imry-Ma (LIM)\,\cite{Larkin70,ImryMa75}
disordered state. 
We shall see that the anisotropy induced by random gauge field
is more robust than that due to random magnetic field.
(This scenario is also relevant
for the two-valley QH in AlAs system\,\cite{Shkolnikov05}.)


The field-induced easy-plane anisotropy 
completely changes 
thermodynamics, transforming an O(3) ferromagnet, which does not order
in 2d, to the XY model which exhibits a  
Berezinskii-Kosterlitz-Thouless
transition to an ordered XY state.
The transition temperature $T_{BKT}$ is logarithmically renormalized by the
out-of-plane fluctuations\,\cite{Hikami80}, 
$T_{BKT} \sim  {J}/{\ln (l_{XY}/\ell_B)}$,
where $l_{XY}$ is the correlation length.
For fields of the order of $30\, {\rm T}$, with $l_{XY}$ 
given by Eq.(\ref{eq:corr_length}) below,
we obtain $T_{BKT}$
in the experimentally accessible range of a few Kelvin. 

The XY-ordered QHFM state hosts fractional $\pm e/2$ charge excitations, 
so-called merons \cite{MacDonald95}. 
Merons are 
vortices such that 
in the vortex core the order parameter smoothly rotates out of the $xy$ plane. 
There are four types of merons \cite{MacDonald95}, since a meron can have 
positive or negative vorticity and the order parameter inside the core can 
tilt either in $+z$ or $-z$ direction. 
A pair of merons with the same charge and opposite vorticity is 
topologically equivalent to a skyrmion
of charge $2(e/2)=e$ \cite{MacDonald95}.
 
Turning to the discussion of QH effect, 
the hierarchy of the spin- and valley-polarized states 
is determined by relative strength of the Zeeman energy
and the randomness-induced anisotropy.
Our estimate below obtains the anisotropy 
of a few Kelvin at $B\sim 30\,{\rm T}$.
This is smaller than the Zeeman energy in graphene, 
$\Delta_z=g\mu_B B \sim 50 \,{\rm K}$ at 
$B \sim 30\, {\rm T}$. Therefore we expect 
that $\nu=0$ state is spin-polarized, 
with both valley states filled. (This was assumed 
in our previous analysis~\cite{Abanin06} of edge states
in $\nu=0$ state.) In contrast, in highly corrugated 
samples, when the anisotropy exceeds the Zeeman energy, 
an easy-plane valley-polarized $\nu=0$ state can be favored.   

While the character of $\nu=0$ state is sensitive to the anisotropy strength,  
the $\nu=\pm1$ states (see Fig.\ref{fig1}a)
are always both spin- and valley-polarized.
Below we focus on $\nu=\pm1$ states, keeping in 
mind that for strong randomnes our discussion also applies
to $\nu=0$ state. 

Zeeman-split free Dirac fermion LL 
are given by
\be\label{LLbulk}
E_n={\rm{sgn}}(n)|2n|^{1/2} \frac{\hbar v}{\ell_B}\pm \Delta_z, \quad \ell_B=
\left(\hbar c/eB \right)^{1/2},
\ee
with $n$ integer and $v\approx 8\times 10^7\, {\rm cm/s}$. 
Each LL is doubly valley-degenerate.
Random field (\ref{eq:delta_h}) couples to electron orbital motion
in the same way as the external field $B$, 
producing a local change in cyclotron energy 
and in the LL density. While the random field 
splits the $n\neq 0$ LL, 
for $n=0$, it does not affect the the single-particle energy (\ref{LLbulk})
and couples to electron dynamics via exchange effects only.
To estimate this coupling, we note 
that the field (\ref{eq:delta_h}) leads 
to valley imbalance in exchange energy per particle:
\be\label{eq:E_exch}
E_{\rm K (K')}=E_{\rm exch}(B\pm\delta h)= \frac{A e^2}
{\kappa\ell_{B\pm \delta h}}, 
\quad A=\left( \frac{\pi}{8} \right)^{1/2}
,
\ee
where $\kappa$ is the 
dielectric constant of graphene.

Let us analyze the graphene QHFM energy dependence on the gauge field.
We consider a fully spin- and valley-polarized $\nu=-1$ state, described by 
a ferromagnetic order parameter $\vec n=(n_1,n_2,n_3)$ in the $K, K'$ valley
space. 
The valley-isotropic exchange interaction gives rise to a sigma model, 
with the gradient term only \cite{MacDonald94}:
\be\label{eq:Jgrad^2}
E_0(n)=\frac 12 \int  J(\nabla\vec n)^2 d^2x, \quad J=\frac{e^2}{64\kappa 
\ell_B}. 
\ee
The valley-asymmetric coupling to $\delta h$ 
in Eq.(\ref{eq:E_exch}) generates a Zeeman-like
Hamiltonian with a uniaxial random field.
\be\label{eq:E1}
E_1(n)=\int g \delta h(x)n_3(x) d^2x
,\quad
g= n\frac{d E_{\rm exch}}{dB}= \frac{An e^2}{2B\ell_{B}}
,
\ee
where $n=1/2\pi \ell _B^2$ is the electron density.

We estimate the energy gain from the order parameter $\vec n(x)$ 
correlations 
with the random field, treating the anisotropy
(\ref{eq:E1}) perturbatively in $\delta h$. 
Decomposing $\vec n(x)=\bar{\vec n} +\delta \vec n(x)$ and taking variation
in $\delta \vec n$,
we obtain
\[
J\nabla^2 \delta \vec n =g \delta \vec h_\perp
,\quad
\delta \vec h_\perp = (\vec z -\bar{\vec n}(\bar{\vec n}\cdot\vec z))\delta h 
.
\]
Substituting the solution for $\delta \vec n$ into the energy functional 
(\ref{eq:Jgrad^2}), (\ref{eq:E1}),
we find 
an energy gain for $\bar{\vec n}$ of the form
\be\label{eq:XYanisotropy}
\delta E=-\lambda \int (1-\bar n^2_3(x)) d^2x
,\quad
\lambda=\sum_k\frac{g^2\la \delta h_{-k}\delta h_k\ra}{2J k^2}
,
\ee
where averaging over spatial fluctuations of $\delta h$ is performed.
This anisotropy favors the $XY$ state, $\bar n_3=0$.
Qualitatively (see Fig.\ref{fig2}), the fluctuations 
due to $\delta{\vec n}$  tilting
towards the $z$-axis minimize the energy of coupling to the uniaxial 
field when $\bar{\vec n}$ is transverse to it.

Now, let us compare the energies of the $XY$ and
the Larkin-Imry-Ma state~\cite{Larkin70, ImryMa75}. 
In LIM state the energy is lowered by domain formation such 
that the order parameter in each domain is aligned with the 
average field in this domain. Polarization varies smoothly between domains, 
and the typical domain size $L$ 
is determined by the balance between domain wall 
and magnetic field energies. 
In our system, the LIM energy per unit area is
\be\label{LIMenergy}
\epsilon _{LIM} \sim -\frac{ g\Phi (L)}{L^2}+\frac{J}{L^2},
\ee
where 
$\Phi (L)$ is typical flux value through 
a region of size $L$. To estimate $\Phi (L)$ we
write the magnetic flux through a region 
of size $L$ as an integral of the vector 
potential over the boundary, which gives 
\be\label{eq:meanflux}
\Phi (L)=\oint a_i (x)\, dx_i  \sim \alpha\sqrt{ L/\xi}.
\ee

Minimizing the LIM energy (\ref{LIMenergy}), 
we find
\be\label{LIMenergy2}
\epsilon _{LIM} \sim - \frac{g^4\alpha ^4}{J^3 \xi ^2}. 
\ee
Comparison to the $XY$ anisotropy $\lambda \sim 
-{g^2\alpha^2}/{J\xi ^2}$ gives 
\be\label{eq:compare}
\frac{\lambda}{\epsilon _{LIM}} \sim \frac{J^2}{ g^2\alpha ^2} \sim 
\left( \frac{\Phi_0}{\Phi(\xi)} \right)^2,
\quad \Phi_0=hc/e,
\ee
where 
the flux through a region of size $\xi$ where
random field does not change sign. 
Interestingly, the ratio (\ref{eq:compare}) does not depend on the external 
magnetic field. 
Therefore, at weak randomness,
when the random 
field flux through 
an area $\xi^2$ is much smaller than the flux quantum, the 
ordered $XY$ state has lower energy
than the disordered LIM state. 

In the opposite limit of strong randomness 
spins align with the local $\delta h$,
forming a disordered state.
It is instructive to note that for
a model with
white noise correlations of magnetic field, rather than of
vector potential, the ratio (\ref{eq:compare}) is of order one. 
In this case the competition of the LIM and the ordered states 
is more delicate.

A different perspective on the random-field-induced ordering 
is provided by analogy with the classical dynamics of 
a pendulum driven at suspension\,\cite{Kapitza}.
The latter, when driven at sufficiently high frequency, 
acquires a steady state 
with the pendulum pointing along the driving force axis.
As discussed in Ref.\,\cite{LandauLifshitz},
this phenomenon can be described by an effective potential $U_{\rm eff}$
obtained by averaging the kinetic energy over fast oscillations,
with the minima of $U_{\rm eff}$ on the driving axis 
and maxima in the equatorial plane
perpendicular to it. 
This behavior is robust upon replacement of periodic driving by 
noise\,\cite{noise-driven-pendulum}.
Our statistical-mechanical problem differs from the pendulum problem 
merely in that the 1d time axis is replaced by
2d position space, which is inessential for the validity of the argument.
The resulting effective potential is thus identical to that
for the pendulum, with the only caveate related to the sign
change $U_{\rm eff}\to -U_{\rm eff}$ in the effective action,
as appropriate for transtion from classical to statistical mechanics.
Thus in our case the minima of $U_{\rm eff}$ are found in the equatorial 
plane, in agreement with the above discussion.

The easy-plane anisotropy (\ref{eq:XYanisotropy}) can be estimated as
\be\label{eq:lambda}
\lambda/n\sim \frac{\delta h^2}{B^2}\times\frac{\xi^2}{\ell_B^2} \times \frac{e^2}{\kappa 
\ell_B}\approx 0.1-10 \ {\rm K/particle},
\ee 
where $\delta h\sim 0.1-1\,{\rm T}$, $\xi\sim 30\,{\rm nm}$ 
and $B\sim 30\,{\rm T}$ was used. 
Since this is smaller than the Zeeman energy, we
expect that the easy-plane 
ferromagnet in the valley space 
is realized at $\nu=\pm 1$, while $\nu=0$ state is spin polarized 
with both valley states filled.  

The out-of-plane fluctuations of the order parameter are characterized by the
correlation length
\be\label{eq:corr_length}
l_{XY}\sim \sqrt{J/\lambda} \approx 1-10\,\ell_B
\ee
for the above parameter values.
The length $l_{XY}$ sets a typical scale 
for order parameter change in the core of vortices
(merons) as well as near edges of the sample and defects
which induce non-zero $z$-component.

To measure the correlation length $l_{XY}$
one may use the spatial structure of $\nu=0$ wavefunction.
Since the $K(K')$ 
electrons reside solely on either $A$ or $B$ sublattice,  
the order parameter $z$-component 
is equal to the density imbalance between 
the two sublattices. The latter can be directly 
measured by STM imaging technique. 

Finally, we briefly outline the calculation of 
QHFM valley anisotropy for a 
pure graphene sheet (the details will be published elsewhere).  
Let us compare the energies of state 1,
in which only the valley $K$ (or $K'$) Landau level is occupied,
and state 2, with electrons in an equal-weight $K$-$K'$ superposition state.
Since the $K\,(K')$ electrons at $\nu=0$ reside
on $A\, (B)$ sublattice, 
the energies per particle in the Hartree-Fock 
approximation are given by
\bea\nonumber
&& 
E_1=\frac{n}{2}\sum _{{\bf r}\in A} v_0 V(r)\left( 1-e^{-r^2/2l_B^2} \right), 
\quad
V(r)=\frac{e^2}{\kappa r},
\\
\nonumber
&&
E_2=\frac{n}{4}\sum_{{\bf r}\in A,B} v_0 V(r)\left( 1-e^{-r^2/2l_B^2} \right),
\eea
where $v_0$ is unit cell volume.
(Here we take $z=0$ to be a site 
of the $A$ sublattice.) 
We approximate the energy difference $E_1-E_2$ by the Fourier harmonic 
of the Hatrtree-Fock energy density 
at the wave vector $\left |{\bf Q}\right | \sim 1/a$, 
where $a\approx 0.14\, {\rm nm}$ is the 
graphene lattice spacing:
 \[
   \Delta E=E_1-E_2\approx \frac{n}{4}\int V(r)\left(1-e^{-r^2/2l_B^2}\right)
   e^{i{\bf Qr}}\, d^2 r. 
 \]
With $Q\ell_B\sim \ell_B/a \gg 1$ at $B\simeq 30\,{\rm T}$, the integral yields
 \be\label{eq:anisotropy}
   \Delta E \approx 
-\frac{27}{512 \pi^3} \times 
  \left(\frac{a}{l_B} \right)^{3} \times \frac{e^2}{\kappa l_B}
\simeq   10\,{\mu} {\rm K}
 \ee
indicating  
that the anisotropy is negligible. 

We note that the situation is completely different for higher LL. 
Goerbig et al.~\cite{Moessner06} pointed out that the Coulomb interaction 
can backscatter electrons of $K$ and $K'$ type at LL with $n\neq 0$, which 
leads to a much stronger lattice anisotropy of the order $a/l_B$. 
This effect is absent for the zeroth LL 
due to the fact that $K$ and $K'$ states
occupy different sublattices.

In summary, we studied the valley symmetry breaking of 
graphene QHFM. We considered the coupling of the strain-induced
random magnetic field and found that it generates an easy-plane anisotropy, 
which is much stronger than the symmetry-breaking terms due to lattice.
The estimates of the 
field-induced anisotropy 
suggest that the random field may be 
a principal mechanism of $K-K'$ QHFM symmetry breaking. 
The easy-plane ordered state is expected to exhibit BKT transition at 
experimentally accessible temperatures and half-integer charge excitations. 

We are grateful to A. K. Geim, M. Kardar and P. Kim for useful discussions.
This work is supported by NSF MRSEC Program (DMR 02132802),
NSF-NIRT DMR-0304019 (DA, LL), and
NSF grant 
DMR-0517222 (PAL).



\begin{references}
\vspace{-4mm}

\bibitem{Novoselov04}
K. S. Novoselov, A. K. Geim, S. V. Morozov, D. Jiang, Y. Zhang, S. V. Dubonos,
I. V. Grigorieva, A. A. Firsov, 
Science, {\bf 306}, 666 (2004);
Proc. Natl. Acad. Sci. U.S.A., {\bf 102}, 10 451 (2005).

\bibitem{Novoselov05}
K. S. Novoselov, A. K. Geim, S. V. Morozov, D. Jiang, M. I. Katsnelson, I. V. Grigorieva, S. V. Dubonos, A. A. Firsov, Nature {\bf 438}, 197 (2005).

\bibitem{Zhang05}
Y. Zhang, Y.-W. Tan, H. L. Stormer and P. Kim, Nature {\bf 438}, 201 (2005). 

\bibitem{Zhang06}
Y. Zhang, Z. Jiang, J. P. Small, M. S. Purewal, Y. W. Tan, M. Fazlollahi, J. D. Chudow, J. A. Jaszaczak, H. L. Stormer, and P. Kim, 
Phys. Rev. Lett., {\bf 96}, 136806 (2006).

\bibitem{QHFM}
For a review, see article by S. M. Girvin and A. H. MacDonald in 
Perspectives in Quantum Hall Effects, S. Das Sarma and A. Pinczuk, 
Eds. (Wiley, New York, 1997).

\bibitem{MacDonald94} K. Yang, K. Moon, L. Zheng, 
 A. H. MacDonald, S. M. Girvin, D. Yoshioka, and S.-C. Zhang, 
Phys. Rev. Lett. {\bf 72}, 732 (1994).

\bibitem{MacDonald95} 
K. Moon, H. Mori, K. Yang, S. M. Girvin, A. H. MacDonald, L. Zheng, 
D. Yoshioka, and S.-C. Zhang, Phys. Rev. B {\bf 51}, 5138 (1995).

\bibitem{Nomura06}
K. Nomura and A. H. MacDonald, Phys. Rev. Lett. {\bf 96}, 256602 (2006)

\bibitem{Moessner06}
M. O. Goerbig, R. Moessner and B. Dou\c{c}ot, cond-mat/0604554, unpublished. 

\bibitem{Alicea06}
J. Alicea and M. P. A. Fisher, 
Phys. Rev. B {\bf 74}, 075422 (2006)

\bibitem{Iordanskii85} 
S. V. Iordanskii and A. E. Koshelev, 
ZETF Pisma, {\bf 41}, 471 (1985)
[translation: JETP Lett. {\bf 41}, 574 (1985)].

\bibitem{KaneMele97} 
C. L. Kane and E. J. Mele, Phys. Rev. Lett. 78, 1932 (1997).

\bibitem{Morpurgo06}
A. Morpurgo and F. Guinea, cond-mat/0603789.

\bibitem{Morozov06}
S. V. Morozov, K. S. Novoselov, M. I. Katsnelson, F. Schedin, 
L. A. Ponomarenko, D. Jiang, and A. K. Geim,
Phys. Rev. Lett. {\bf 97}, 016801 (2006). 

\bibitem{Aharony78} A. Aharony, Phys. Rev. B{\bf 18}, 3328 (1978).

\bibitem{Feldman98} D. E. Feldman, J. Phys. A: Math. Gen. {\bf 31}, 177 (1998).

\bibitem{Pelcovits85} B. J. Minchau and R. A. Pelcovits,
Phys. Rev. B{\bf 32}, 3081 (1985).

\bibitem{Wehr06} 
J. Wehr, A. Niederberger, L. Sanchez-Palencia and M. Lewenstein, 
cond-mat/0604063.

\bibitem{Larkin70} 
A. I. Larkin, 
JETP {\bf 31}, 784 (1970).

\bibitem{ImryMa75} 
Y. Imry and S. K. Ma, 
Phys. Rev. Lett. {\bf 35}, 1399 (1975).



\bibitem{Shkolnikov05}
Y. P. Shkolnikov, S. Misra, N. C. Bishop, E. P. De Poortere, and M. Shayegan,
 Phys. Rev. Lett. {\bf 95}, 066809 (2005) 

\bibitem{Hikami80} 
S. Hikami, T. Tsuneto, Progr. Theor. Phys. {\bf 63}, 387 (1980).

\bibitem{Abanin06}
D. A. Abanin, P. A. Lee and L. S. Levitov, 
Phys. Rev. Lett. 96, 176803 (2006).

\bibitem{Kapitza} P. L. Kapitza, Zh. Eksp. Teor. Fiz., {\bf 21}, 588 (1951).

\bibitem{LandauLifshitz} L. D. Landau and E. M. Lifshitz, Mechanics,
Chap. V, \S 30,
(Reed Educational and Pofessional Publ. Ltd, 2002)

\bibitem{noise-driven-pendulum} R. L. Stratonovich and Yu. M. Romanovsky,
Nauch. Dokl. Vys. Shk., ser. fiz.-mat. {\bf 3}, 221 (1958) (in Russian).


























 

\end{references}
\end{document}